\documentclass[aps,prl,twocolumn,showpacs,showkeys,footinbib,amsmath,amssymb,superscriptaddress]{revtex4-2}

\usepackage{amsmath}
\usepackage{amssymb}
\usepackage{graphicx}
\usepackage{bm}
\usepackage{color}
\usepackage{relsize}
\usepackage{braket}
\usepackage{bbold}
\usepackage{txfonts}
\usepackage{epstopdf}
\usepackage{hyperref}

\begin{document}
	
\title{Emergent Multiferroic Altermagnets and Spin Control via Noncollinear Molecular Polarization}
	
\author{Ziye Zhu}
\affiliation{Eastern Institute for Advanced Study, Eastern Institute of Technology, Ningbo, Zhejiang 315200, China}
\affiliation{Department of Physics, University of Science and Technology of China, Hefei, 230026, China}
	
\author{Yuntian Liu}
\affiliation{Department of Physics, University at Buffalo, State University of New York, Buffalo, New York 14260, USA}
	
\author{Xunkai Duan}
\affiliation{Eastern Institute for Advanced Study, Eastern Institute of Technology, Ningbo, Zhejiang 315200, China}
\affiliation{School of Physics and Astronomy, Shanghai Jiao Tong University, Shanghai 200240, China}
	
\author{Jiayong Zhang}
\affiliation{Eastern Institute for Advanced Study, Eastern Institute of Technology, Ningbo, Zhejiang 315200, China}
\affiliation{Department of Physics, University of Science and Technology of China, Hefei, 230026, China}
\affiliation{School of Physical Science and Technology, Suzhou University of Science and Technology, Suzhou, 215009, China}

\author{Bowen Hao}
\affiliation{Eastern Institute for Advanced Study, Eastern Institute of Technology, Ningbo, Zhejiang 315200, China}
	
\author{Su-Huai Wei}
\affiliation{Eastern Institute for Advanced Study, Eastern Institute of Technology, Ningbo, Zhejiang 315200, China}
	
\author{Igor \v{Z}uti\'c}
\affiliation{Department of Physics, University at Buffalo, State University of New York, Buffalo, New York 14260, USA}
	
\author{Tong Zhou}
\email{tzhou@eitech.edu.cn}
\affiliation{Eastern Institute for Advanced Study, Eastern Institute of Technology, Ningbo, Zhejiang 315200, China}
	
\date{\today}

\vspace{1em}

\begin{abstract}

Altermagnets, with spin splitting and vanishing magnetization, have been attributed to many fascinating phenomena and potential applications. In particular, integrating ferroelectricity with altermagnetism to enable magnetoelectric coupling and electric control of spin has drawn significant attention. However, its experimental realization and precise spin manipulation remain elusive. Here, by focusing on molecular ferroelectrics, the first discovered ferroelectrics renowned for their highly controllable molecular polarizations and structural flexibility, we reveal that these obstacles can be removed by an emergent multiferroic altermagnets with tunable spin polarization in a large class of fabricated organic materials. Using a symmetry-based design and a tight-binding model, we uncover the underlying mechanism of such molecular ferroelectric altermagnets and demonstrate how noncollinear molecular polarization can switch the spin polarization on and off and even reverse its sign, as detectable by the magneto-optical Kerr effect. From the first-principles calculations, we verify the feasibility of these materials in a series of well-established hybrid organic-inorganic perovskites and metal-organic frameworks. Our findings bridge molecular ferroelectrics and altermagnetic spintronics, highlighting an unexplored potential of multifunctional organic multiferroics.
\end{abstract}

\maketitle

Electrical control of magnetism is a fundamental challenge in condensed matter physics, with significant implications for high-density data storage and energy-efficient spintronic devices~\cite{zutic2004:RMP,Tsymbal:2019,Ohno2000:Nature}. Magnetoelectric coupling, which connects magnetic and electric ordering, is crucial in multiferroics~\cite{Eerenstein2006,Cheong2007,Dong2015,Fiebig2016:evolution}. Traditionally, multiferroics leverage antiferromagnets due to their insulating nature, absence of stray fields, and ultrafast dynamics~\cite{Baltz2018,Smejkal2018,Chen2024,Shao2024:npjspintronics}. However, the lack of spin polarization in antiferromagnets limits their potential in spintronics. The growing interest in altermagnets (AM)~\cite{Smejkal2022a,Mazin2024:altermagnetism,wright2025altermagnets,hayami2019momentum,
	yuan2020giant,ma2021multifunctional,mazin2021prediction,
	Bai2024,song2025altermagnets,Wei2024,Tamang2024,Fender2025,Liu2025,Krempasky2024:MnTe,Amin2024,
	SongAMNature2025,jiang2025metallic,zhang2025crystal,Denisov2024,guo2025spin, CheJACS2025, PhysRevLett.134.136301, cao2025symmetry}, which unify the spin polarization of ferromagnets (FM) and the zero net magnetization of antiferromagnets (AFM), significantly broadens the avenues for realizing magnetoelectric coupling in multifunctional materials.

Integrating (anti)ferroelectricity with altermagnetism offers a new route toward multiferroics. Recent discoveries about altermagnetoelectric coupling~\cite{Duan2025PRL,zhu2025two,Gu2025PRL, Smejkal2024arxiv,Sun2024a, guo2025altermagnetic}, enable electric control of magnets, highlighting the potential of AM for developing electrically controlled spintronic and multiferroic devices. The key to such integration is the ability of (anti)ferroelectricity to connect opposite-spin sublattices via rotation, $R$, symmetry rather than translation, $t$, or inversion, $I$, symmetries. Several strategies, including antiferroelectricity~\cite{Duan2025PRL}, lattice distortion~\cite{zhu2025two,Gu2025PRL, Smejkal2024arxiv}, and layer sliding/stacking~\cite{Sun2024a, sheng2024ubiquitous, zhu2025sliding, sun2025proposing} have been explored to realize this symmetry-driven coupling, yet experimental realization and precise spin control remain elusive.

Molecular ferroelectricity, as the first ferroelectricity discovered in 1921~\cite{Valasek1921}, presents a versatile alternative~\cite{Wang2023a} for realizing multiferroic altermagnets. Unlike conventional ferroelectrics, molecular ferroelectrics derive their polarization from the alignment of molecular dipoles, providing exceptional structural tunability, mechanical flexibility, and multifunctionality~\cite{Pan2024}. The ability to precisely engineer molecular size, shape, charge, and polarity enables versatile control over molecular polarization, $\mathbf{\mathit{P}_M}$~\cite{Pan2024,Zhang2019}, which has been experimentally realized using external stimuli such as electric field~\cite{Pan2024,Zhang2019,xu2019hybrid,Liu2022,Su2023,Di2013,Jain2016, Lou2020}, pressure~\cite{Liu2022,Su2023}, temperature~\cite{Liu2022,Su2023}, and light~\cite{Ren2025}. Moreover, the weak spin-orbit coupling (SOC) and hyperfine interactions, intrinsic to light elements such as C, H, O, and N, contribute to excellent spin properties, including long spin-diffusion lengths and relaxation times~\cite{Wang2023a, Liu2022, Dediu2009, Li2021}. Despite these advantages, integrating AM with molecular ferroelectricity remains unexplored, representing a significant gap in the advancement of molecular multiferroics and their devices.

\begin{figure*}[t!]
	\centering
	\vspace{0.2cm}
	\includegraphics*[width=0.75\textwidth]{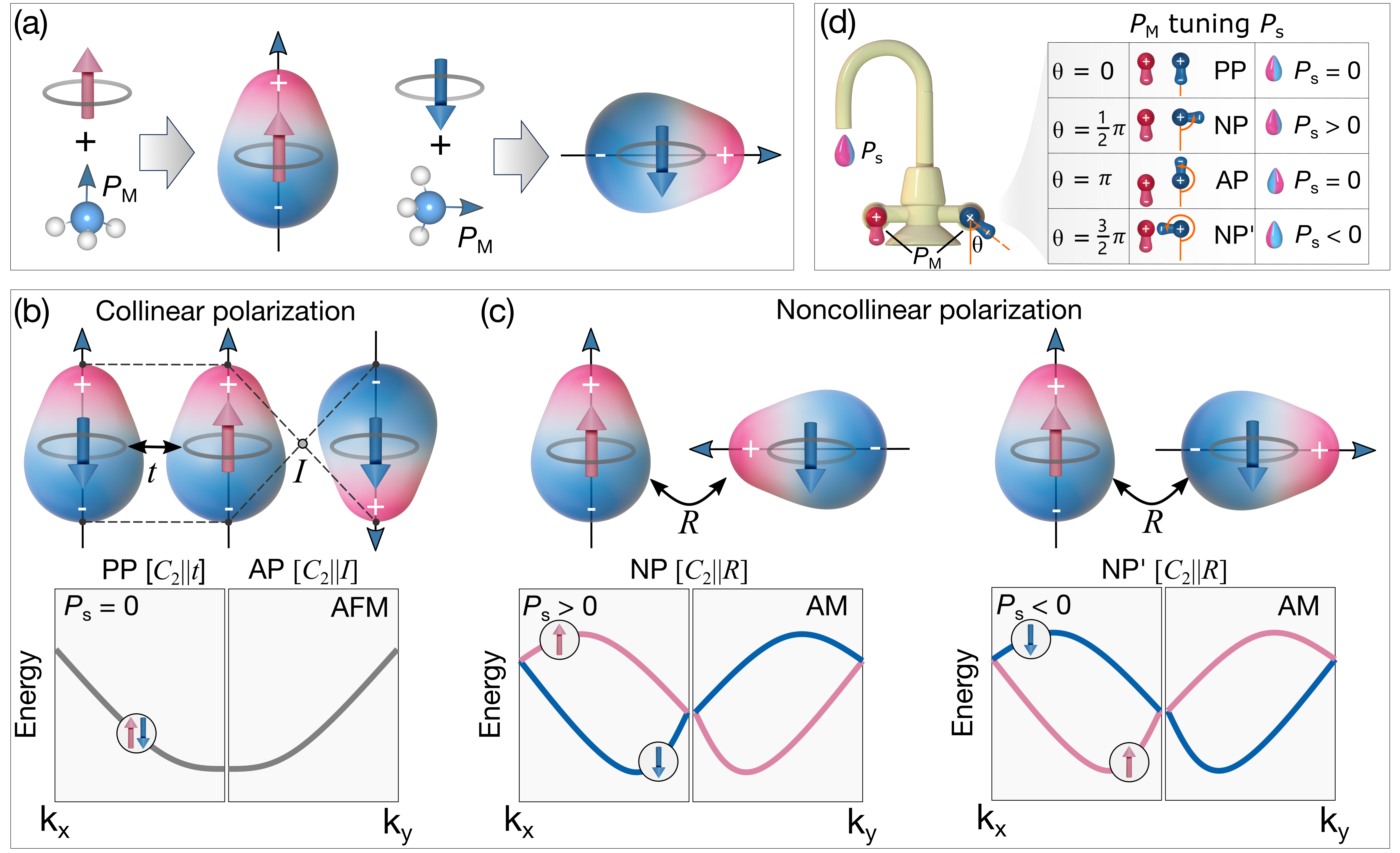}
	\vspace{-0.2cm}
	\caption{ \textbf{Design principle of MFEAM with spin control via $\mathbf{\mathit{P}_M}$ twist.} {\bf (a)} The fundamental building block consists of magnetic atoms (arrows with circles) and polar molecules with $\mathbf{\mathit{P}_M}$, directed from the negative to the positive charge center. {\bf (b)} Parallel (PP) and antiparallel (AP) $\mathbf{\mathit{P}_M}$ establish magnetic sublattices related by $t$ and $I$ symmetry, resulting in conventional AFM. {\bf (c)} Noncollinear (NP) $\mathbf{\mathit{P}_M}$ configurations connect magnetic sublattices via $R$ symmetry, giving rise to AM with finite $\mathbf{\mathit{P}_S}$ (assumed $\mathbf{\mathit{P}_S}>0$ for clarity). Reversing one $\mathbf{\mathit{P}_M}$ in the NP configuration (denoted as NP$^\prime$) flips $\mathbf{\mathit{P}_S}$. {\bf (d)} Schematic illustration of tuning $\mathbf{\mathit{P}_S}$ through $\mathbf{\mathit{P}_M}$, where $\mathbf{\mathit{P}_M}$ is indicated by the faucet handle, and $\mathbf{\mathit{P}_S}$ by water drops with red (blue) for spin-up (-down). Twisting $\mathbf{\mathit{P}_M}$ enables the switching $\mathbf{\mathit{P}_S}$ on/off and reversing its sign.}
	\label{Figure1}
\end{figure*}

In this work, we present a universal design framework for molecular ferroelectric altermagnets (MFEAM), which combines the momentum-dependent spin splitting of AM with the structural flexibility of molecular ferroelectricity. Using spin-group theory~\cite{litvin1974spin,liu2022spin,Spingroup_SongPRX,Spingroup_FangPRX,Spingroup_LiuPRX,Spingroup_LiuMagnon}, we identify the critical role of noncollinear $\mathbf{\mathit{P}_M}$ in creating $\it{R}$ symmetry, connecting spin sublattices without $\it{t}$ or $\it{I}$ symmetries. Molecular systems inherently support this symmetry through their controllable polarizability and spatial flexibility. Employing a tight-binding (TB) model, we show that the $\mathbf{\mathit{P}_M}$ distribution governs the magnetic properties, and that twisting $\mathbf{\mathit{P}_M}$ can effectively toggle spin polarization, $\mathbf{\mathit{P}_S}$, on/off and even reverse its sign. Guided by this theoretical framework, we use first-principles calculations to propose a series of promising MFEAM candidates, ranging from hybrid organic-inorganic perovskites (HOIPs) to metal-organic frameworks (MOFs). These findings lay the groundwork for advancing molecular multiferroics and developing the next-generation multifunctional devices.

\vspace{1em}
\indent{$Design~principle$}---
Our proposed framework for MFEAM is illustrated in Figure~\ref{Figure1}, where the basic unit consists of magnetic atoms forming AFM order and their surrounding molecules with $\mathbf{\mathit{P}_M}$. When $\mathbf{\mathit{P}_M}$ are collinear, either parallel (PP) or antiparallel (AP), the resulting state is a conventional AFM characterized by vanishing $\mathbf{\mathit{P}_S}$ and spin-group symmetries $\left[C_2 \|t\right]$ or $\left[C_2 \|I\right]$~\cite{Smejkal2022a, Bai2024} as shown in [Figure~\ref{Figure1}(b)]. Here, $C_2$ denotes the two-fold rotation operator in spin space, while $t$ and $I$ represent translation and inversion operations in crystal structures, respectively. In contrast, when $\mathbf{\mathit{P}_M}$ become noncollinear (NP), an AM state emerges, featuring finite $\mathbf{\mathit{P}_S}$ and symmetry characterized by $\left[C_2 \|R\right]$ [Figure~\ref{Figure1}(c)]. Notably, reversing a single $\mathbf{\mathit{P}_M}$ flips the $\mathbf{\mathit{P}_S}$ sign. Thus, by simply twisting $\mathbf{\mathit{P}_M}$ in MFEAM, one can tune its magnetism, control the presence of $\mathbf{\mathit{P}_S}$, and reverse its sign, all without changing the magnetic order. Since the $\mathbf{\mathit{P}_M}$ can be twisted by electric field~\cite{Pan2024,Zhang2019,xu2019hybrid,Liu2022,Su2023,Di2013,Jain2016, Lou2020}, temperature~\cite{Liu2022,Su2023}, and light~\cite{Ren2025}, this tunable behavior enables promising spintronic and multiferroic functionalities [Figure~\ref{Figure1}(d)].

\begin{figure*}[t!]
	\centering
	\includegraphics[width=0.98\textwidth]{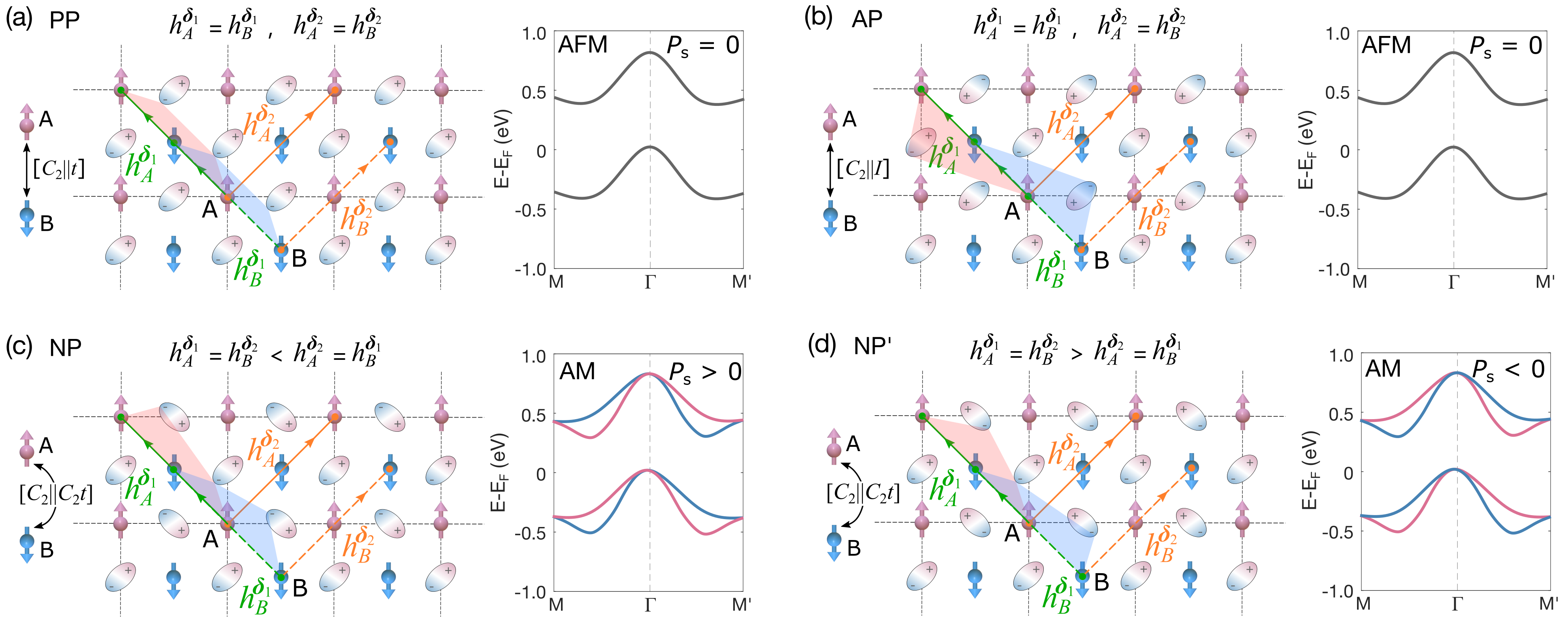}
	\caption{ \textbf{Effective model.} Schematic of 2D square magnetic sublattices A and B surrounded by polar molecules (ellipses), forming PP, AP, NP, and NP’ configurations with spin lattices connected by symmetry $t$, $I$, $C_2t$, $C_2t$, respectively. The solid (dashed) arrows indicate 3NN hopping vectors ($\delta_{1,2}$) in A (B) sublattice, where the red (blue) shades illustrate how its hopping strengths $h_A^{{\textbf{$\delta$}_{1,2}}}$ ($h_B^{{\textbf{$\delta$}_{1,2}}}$) are affected by $\mathbf{\mathit{P}_M}$ distribution. The other hopping vectors $\delta_{3,4}$ are not shown as their hopping strengths along opposite directions are identical ($h_{A,B}^{{\textbf{$\delta$}_{3,4}}}$ = $h_{A,B}^{{\textbf{$\delta$}_{1,2}}}$). Spin-resolved bands are displayed in black (spin-degenerate), red (spin-up), and blue (spin-down), calculated using the TB model. The critical 3NN parameters are {\bf (a)} and {\bf (b)} $h_{A,B}^{{\textbf{$\delta$}{_{1,3}}}}$ = 0.03 eV, $h_{A,B}^{{\textbf{$\delta$}{_{2,4}}}}$ = 0.08 eV; {\bf (c)} $h_{A}^{{\textbf{$\delta$}{_{1,3}}}}$ = $h_{B}^{{\textbf{$\delta$}{_{2,4}}}}$ = 0.03 eV, $h_{A}^{{\textbf{$\delta$}{_{2,4}}}}$ = $h_{B}^{{\textbf{$\delta$}{_{1,3}}}}$ = 0.08 eV; {\bf (d)} $h_{A}^{{\textbf{$\delta$}{_{1,3}}}}$ = $h_{B}^{{\textbf{$\delta$}{_{2,4}}}}$ = 0.08 eV, $h_{A}^{{\textbf{$\delta$}{_{2,4}}}}$ = $h_{B}^{{\textbf{$\delta$}{_{1,3}}}}$ = 0.03 eV, with $M_A$ =-$M_B$ = 0.4 eV. All other parameters and model details including high-symmetry $k$-points are provided in Supplementary Note S1.}
	\label{Figure2}
\end{figure*} 

\vspace{1em}
\indent{$Effective~model$}---
To elucidate the mechanism of our proposed MFEAM, we construct an effective tight-binding (TB) model on a general 2D square lattice with nested AFM order, as illustrated in Figure~\ref{Figure2}. This setup incorporates the essential components of our proposal: AFM sublattices with collinear (noncollinear) $\mathbf{\mathit{P}_M}$ connected by $\it{t}$/$\it{I}$ ($\it{R}$) symmetry. Our symmetry analysis has revealed that the emergence of AM (AFM) depends on the inequivalence (equivalence) of the magnetic sublattices governed by the $\mathbf{\mathit{P}_M}$ distribution. To accurately capture such $\mathbf{\mathit{P}_M}$ influence, the model necessitates hopping terms up to the third-nearest neighbor. Thus, the minimal TB model for MFEAM is 

\begin{equation}
	\begin{aligned}
		H &= \left( \sum_{i, j} \left( f_i^{\textbf{$\eta$}_j} c_i^\dagger c_{i+\textbf{$\eta$}_j} + g_i^{\textbf{$\kappa$}_j} c_i^\dagger c_{i+\textbf{$\kappa$}_j} + h_i^{\textbf{$\delta$}_j} c_i^\dagger c_{i+\textbf{$\delta$}_j} \right) + \text{H.C.} \right) \\&+ M_{A,B}\sum_{i\in A,B}c_i^\dagger \sigma_{z} c_{i}.
	\end{aligned}
\end{equation}

\noindent Here, $c_i^\dagger$ and $c_i$ represent electron creation and annihilation operators at site $\it{i}$. The parameters $f_i^{{\textbf{$\eta$}_j}}$, $g_i^{{\textbf{$\kappa$}_j}}$, and $h_i^{{\textbf{$\delta$}{_j}}}$ describe the hopping between site $i$ and its first (NN), second (2NN), and third (3NN) nearest-neighbors, which are connected by the vectors of $\textbf{$\eta$}_j$, $\textbf{$\kappa$}_j$, and $\textbf{$\delta$}_j$, respectively. $M_{A,B}$ denotes the AFM on-site exchange field on sublattices $A$ and $B$.

According to our TB model analysis, Neither NN nor 2NN hoppings contribute to AM as they fail to induce the necessary sublattice inequivalence regardless of the $\mathbf{\mathit{P}_M}$ distribution, as discussed in the Supplementary Note S1. In contrast, 3NN hoppings between sublattices can be either equivalent or inequivalent depending on the $\mathbf{\mathit{P}_M}$ configuration. Specifically, in the PP and AP configurations, the system retains the symmetries $\left[C_2 \|t\right]$ and $\left[C_2 \|I\right]$, ensuring sublattice-equivalent 3NN hoppings ($h_A^{{\textbf{$\delta$}_{1-4}}}$ = $h_B^{{\textbf{$\delta$}_{1-4}}}$). This leads to conventional AFM, as confirmed by the spin-degenerate bands in Figure~\ref{Figure2}(a) and Figure~\ref{Figure2}(b). However, the situation changes completely in the NP configuration. The noncollinear $\mathbf{\mathit{P}_M}$ breaks both $t$ and $I$ symmetries, leaving the sublattices connected by a rotation symmetry, $C_2t$. This leads to inequivalent 3NN hoppings, with $h_A^{{\textbf{$\delta$}_{1,3}}}$ = $h_B^{{\textbf{$\delta$}_{2,4}}}$ $\textless$ $h_A^{{\textbf{$\delta$}_{2,4}}}$ = $h_B^{{\textbf{$\delta$}_{1,3}}}$, and gives rise to the desired MFEAM, as evidenced by spin-polarized bands [Figure~\ref{Figure2}(c)]. Remarkably, reversing a single $\mathbf{\mathit{P}_M}$ to form the NP$^\prime$ configuration interchanges the 3NN hopping parameters ($h_A^{{\textbf{$\delta$}_{1,3}}}$ = $h_B^{{\textbf{$\delta$}_{2,4}}}$ $\textgreater$ $h_A^{{\textbf{$\delta$}_{2,4}}}$ = $h_B^{{\textbf{$\delta$}_{1,3}}}$), thereby producing an AM state with reversed $\mathbf{\mathit{P}_S}$ compared to that in the NP configuration [Figure~\ref{Figure2}(d)].

These TB model results clearly reveal the underlying mechanism of our MFEAM design, demonstrating how $\mathbf{\mathit{P}_M}$ configurations govern magnetism and spin polarization. Since all above four configurations can be realized simply by twisting $\mathbf{\mathit{P}_M}$, this provides a compelling approach for toggling $\mathbf{\mathit{P}_S}$ on and off or reversing its sign without changing the magnetic order. This capability holds great promise for spintronic applications and nonvolatile multiferroic devices~\cite{zutic2004:RMP,Tsymbal:2019,Ohno2000:Nature,Eerenstein2006,Cheong2007,Dong2015,Fiebig2016:evolution}.

\vspace{1em}
\indent{$Material~Realization$}---
Our design principle offers a universal framework for realizing MFEAM based on symmetry analysis, which has been further extended into a Hamiltonian description via our effective model. To bridge theory and experiment, the key criteria for potential candidates are the presence of AFM order and noncollinear $\mathbf{\mathit{P}_M}$ distributions. A promising strategy is to start with AFM materials and incorporate polar molecules to induce the necessary $\mathbf{\mathit{P}_M}$ configurations, which can be selectively controlled during fabrication and tuned through external stimuli such as electric fields, pressure, and light~\cite{Pan2024,Zhang2019,xu2019hybrid,Liu2022,Su2023,Di2013,Jain2016, Lou2020, Ren2025}.

\begin{figure*}[t!]
	\centering
	\includegraphics*[width=0.93\textwidth]{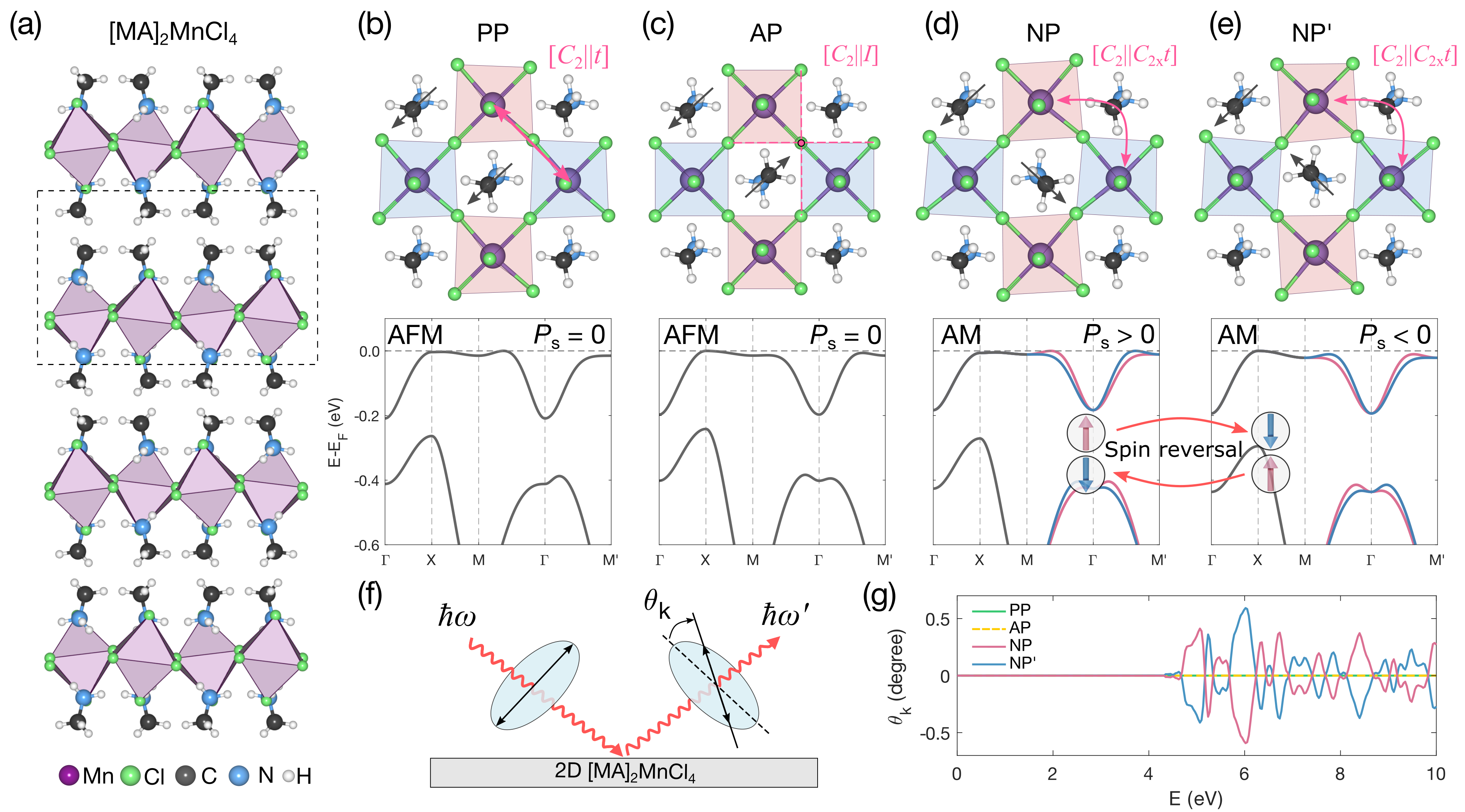}
	\caption{ \textbf{MFEAM realization in [MA]$_2$MnCl$_4$.} {\bf (a)} Crystal structure of layered [MA]$_2$MnCl$_4$, with the monolayer unit highlighted by a dashed box. {\bf (b)}–{\bf (e)} Top views of the [MA]$_2$MnCl$_4$ monolayer in the PP, AP, NP, and NP$^\prime$ configurations, along with the corresponding calculated spin-resolved electronic band structures. {\bf (f)} Schematic illustration of the magneto-optical Kerr effect measurement setup. {\bf (g)} Calculated MOKE signals for each configuration, characterized by the Kerr rotation angle $\theta_k$.} 
	\label{Figure3}
\end{figure*}

\begin{figure*}[t!]
	\centering
	\includegraphics[width=0.93\textwidth]{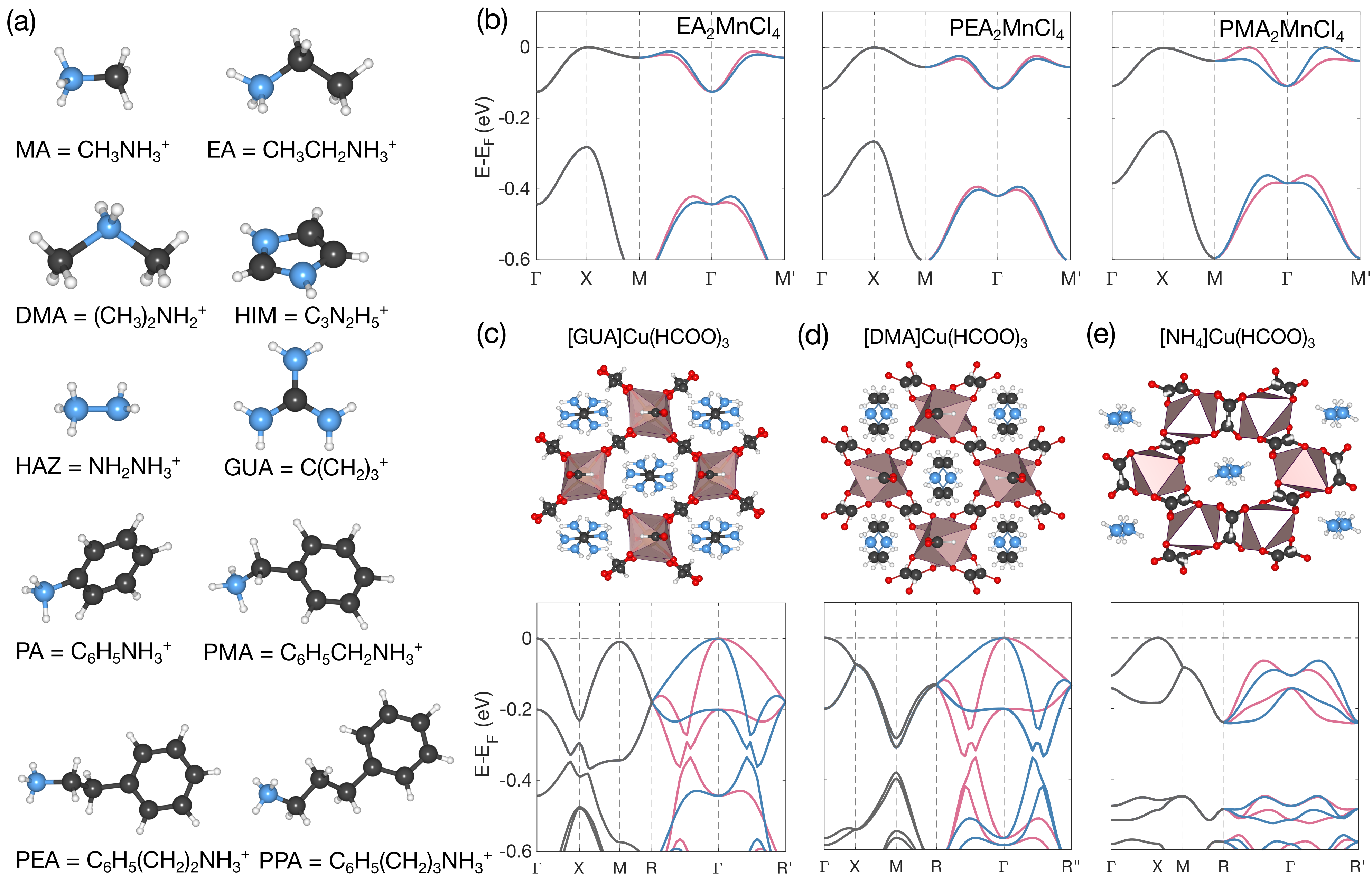}
	\caption{ \textbf{MFEAM candidates in HOIPs and MOFs.} {\bf (a)} Structural diversity of organic polar molecules. {\bf (b)} Calculated band structures of Mn-based HOIP for MFEAM, including [EA]$_2$MnCl$_4$, [PEA]$_2$MnCl$_4$ and [PMA]$_2$MnCl$_4$. {\bf (c)}–{\bf (e)} Crystal structures and corresponding bands of MOFs for MFEAM, including [GUA]Cu(HCOO)$_3$, [DMA]Cu(HCOO)$_3$, [NH$_4$]Cu(HCOO)$_3$.}
	\label{Figure4}
\end{figure*}

Here, we focus on two well-established families of magnetic organic materials: HOIPs and MOFs. These materials are widely studied in energy applications, catalysis, spintronics, and multiferroics, owing to their facile synthesis, structural diversity, and cost-effectiveness~\cite{Liu2022,Li2017,Lu2024}. We first examine magnetic HOIPs with the general formula A$_2$MX$_4$, where A epresents an organic molecular cation, M is a divalent transition metal ion that governs the magnetic properties, and X is a halogen anion. The magnetism in HOIPs is primarily dominated by M, for instance, Mn commonly induces AFM order~\cite{Septiany2021}, while its $\mathbf{\mathit{P}_M}$ distribution mainly depends on A~\cite{xu2019hybrid}. Notably, all our proposed $\mathbf{\mathit{P}_M}$ configurations, PP, AP, and NP/NP$^\prime$ with AFM order have been achieved experimentally through appropriate selection of A, M, and X, combined with controlled synthesis conditions~\cite{Pan2024,xu2019hybrid,Zhang2019,Liu2022,Su2023}.

As a representative example, we consider layered van der Waals material [MA]$_2$MnCl$_4$ (where MA = CH$_3$NH$_3$, methylammonium) due to its relatively simple crystal structure [Figure~\ref{Figure3}(a)] and its experimentally observed AFM ground state~\cite{Lee2000,Kim2021}. Our first-principles calculations focus on the ``organic–inorganic–organic'' 2D monolayer, each monolayer consists of corner-sharing [MnCl$_6$] octahedra surrounded by MA molecules, which dictate the properties of [MA]$_2$MnCl$_4$. The $\mathbf{\mathit{P}_M}$ distribution is controlled by the orientation of the MA molecules. Here, we constructed four different configurations corresponding to our design principle, as shown in Figure~\ref{Figure3}(b)-(e). A more detailed discussion of additional configurations can be found in Supplementary Note S2. Experimentally, HOIP materials with various molecular orientations have already been fabricated and successfully controlled using external electric fields, pressure, and light~\cite{Pan2024,Zhang2019,xu2019hybrid,Liu2022,Su2023,Di2013,Jain2016, Lou2020, Ren2025}, providing strong experimental support for the realization and control of MFEAM.

PP configuration belongs to the $Pmm2$ space group, where the presence of $t$ symmetry restores conventional AFM [Figure~\ref{Figure3}(b)]. Similarly, in the AP configuration, the system adopts the $Cmmm$ space group, where its $I$ symmetry enforces AFM [Figure~\ref{Figure3}(c)]. These symmetry-dependent magnets can be further understood through the $\mathbf{\mathit{P}_M}$-induced distortions. The presence of $\mathbf{\mathit{P}_M}$ affects the tilt of [MnCl$_6$] octahedra, modifying the charge distribution and altering symmetry between spin sublattices, ultimately determining the $\mathbf{\mathit{P}_S}$. Conversely, when the MA molecules align into a NP configuration, the system transitions to the $Pmc2_1$ space group, where a $C_{2x}t$ symmetry connects the magnetic sublattices, inducing AM, as evidenced by the calculated band structure showing a finite $\mathbf{\mathit{P}_S}$ [Figure~\ref{Figure3}(d)]. Furthermore, flipping a single MA molecule to form the NP$^\prime$ configuration reverses the direction of $\mathbf{\mathit{P}_S}$ [Figure~\ref{Figure3}(e)], fully consistent with our TB model. The detailed symmetry analysis of different configurations is provided in the Supplementary Note S3, confirming that the variation in symmetry operations induced by the $\mathbf{\mathit{P}_M}$ distribution plays a decisive role in the emergence of AM.

Since [MA]$_2$MnCl$_4$ is composed primarily of light elements, spin-orbit coupling is weak, as confirmed by our calculations in Supplementary Note S4, facilitating experimental identification of the altermagnetic spin splitting via ARPES, nanoscale imaging, or spin transport~\cite{Krempasky2024:MnTe,Amin2024,SongAMNature2025}. Here, we propose probing the characteristics of altermagnetism through the detection of the magneto-optical Kerr effect (MOKE)~\cite{fan2017electric,yang2020magneto,ding2023magneto,sun2024stacking}. As shown in Figure~\ref{Figure3}(g), the PP and AP configurations exhibit negligible Kerr signals. In contrast, a pronounced Kerr response is observed in the NP/NP$^\prime$ configurations, which can be attributed to the anisotropic optical conductivity induced by the breaking of time-reversal symmetry. It is noteworthy that the Kerr angle magnitude serves as an indicator of spin splitting strength, as reported in previous studies~\cite{OPPENEER2001229, sunadv2025}. Furthermore, the Kerr angle exhibits a sign reversal between the NP and NP$^\prime$ configurations, consistent with the reversal of spin splitting observed in the band structures. The demonstrated $\bf {P_M}$ control of the Kerr response further highlights the robustness of magnetoelectric coupling in the proposed MFEAM.

Our MFEAM design can be extended by substituting MA in [MA]$_2$MnCl$_4$ with other common organic cations such as ethylammonium (EA), dimethylammonium (DMA), and phenylammonium (PA), as illustrated in Figure~\ref{Figure4}(a). These molecules preserve the Mn-based AFM sublattice and tend to adopt NP configurations, making them promising MFEAM candidates, as confirmed by our calculations [Figure~\ref{Figure4}(b) and Supplementary Note S5]. Notably, replacing MA with PMA in [PMA]$_2$MnCl$_4$ enhances the spin splitting from 26 meV to 41 meV due to the larger molecular size, which induces stronger distortions in the [MnCl$_6$] octahedra. This effect provides a practical means to enhance $\mathbf{\mathit{P}_S}$, making altermagnetic behavior more prominent.

Additionally, replacing the Cl component in A$_2$MnCl$_4$ HOIPs with organic anions including [GUA]Cu(HCOO)$_3$, [DMA]Cu(HCOO)$_3$, [NH$_4$]Cu(HCOO)$_3$ leads to the formation of MOFs, offering another pathway to expand MFEAM candidates as verified in our calculations [Figures~\ref{Figure4}(c)–(e)]. Such large organic anions can introduce stronger structural distortions, further enhancing the altermagnetic spin splitting, which reaches 170 meV in [DMA]Cu(HCOO)$_3$, strong enough to support spintronic applications. Although this work focuses on organic materials, our MFEAM design principle can be extended to inorganic systems. For instance, we find the reported multiferroic BaFe$_2$Se$_3$~\cite{Orlandi2014} and Pb$_2$MnWO$_6$~\cite{Dong2014} are inorganic MFEAM, as discussed in Supplementary Note S5.

\vspace{1em}
Our proposed MFEAM bridges two previously separate concepts, molecular ferroelectricity and altermagnetism, into a single material platform, establishing a universal framework for expanding organic multiferroics and realizing novel magnetoelectric coupling. Many of our proposed MFEAM candidates are composed primarily of light elements, making them ideal for studying organic spin dynamics dominated by nonrelativistic spin splitting rather than SOC effects~\cite{bliokh2015spin, Amundsen2024}. Moreover, the $\mathbf{\mathit{P}_S}$ in MFEAM can be flexibly controlled by twisting $\mathbf{\mathit{P}_M}$ via electric fields, pressure, or light, enabling efficient, zero-magnetic-field spin manipulation. This versatile control paves the way for developing electrically or optically driven spintronic devices. In addition, several of our proposed MFEAM materials are already recognized for their advanced performance in solar energy and catalysis~\cite{wright2012organic, Shi2017Lead, Jiao2018Metal}. The incorporation of these newly identified multiferroic properties further broadens their scope, positioning them as highly promising candidates for multifunctional applications~\cite{Liang2023,Cui2025,Zhou2023NM}.

\vspace{1em}
\indent{$Acknowledgements$}---
This work is supported by the National Natural Science Foundation of China (12474155, 12447163, and 11904250), the Zhejiang Provincial Natural Science Foundation of China (LR25A040001), the China Postdoctoral Science Foundation (2025M773440), and the U.S. DOE, Office of Science BES, Award No. DE-SC0004890 (Y.L, I.\v{Z}.). The computational resources for this research were provided by the High Performance Computing Platform at the Eastern Institute of Technology, Ningbo. Z.Z. and Y.L. contributed equally to this work.

\bibliography{referenceV2}

\end{document}